\documentclass[a4paper,11pt]{article}

\usepackage{contribution}

\newcommand{\weblink}[2][]{%
    \ifthenelse{\equal{#1}{}}%
    {\textnormal{\url{#2}}}%
    {\textnormal{\href{#2}{#1}}}%
}

\def\beq{\begin{equation}}
\def\eeq#1{\label{#1}\end{equation}}
\def\eeqn{\end{equation}}

\def\beqa{\begin{eqnarray}}
\def\eeqa#1{\label{#1}\end{eqnarray}}
\def\eeqan{\end{eqnarray}}

\let\bar=\overbar

\def\Dslash{\not{\hbox{\kern-4pt $D$}}}
\def\dslash{\not{\hbox{\kern-2pt $\del$}}}

\def\msb{{\bar{\ssstyle M \kern -1pt S}}}

\newcommand{\contribution}[7][]{%
  \clearpage
  \thispagestyle{plain}
  \ifthenelse{\equal{#1}{}}
  {\hypersetup{pdftitle={#2}}}
  {\hypersetup{pdftitle={#1}}}
  \hypersetup{pdfauthor={{#3} {#4}}}
  {\centering\normalfont\LARGE\bfseries\sffamily #2 \par\nobreak}
  \lhead{}
  \chead{%
    \textit{\footnotesize XIV International Conference on Hadron Spectroscopy
      (\weblink[\textit{hadron2011}]{http://www.hadron2011.de}), 13-17 June 2011, Munich, Germany}%
  }
  \rhead{}
  \bigskip
  \begin{center}
    {#3} {#4}\ifthenelse{\equal{#6}{}}{}{\footnote{\weblink[#6]{mailto:#6}}}
    \ifthenelse{\equal{#7}{}}{}{#7} \\
    \textit{#5}
  \end{center}
  \bigskip
}

\renewcommand{\abstract}[1]{%
  \begin{center}
    \begin{minipage}{0.85\textwidth}
      \begin{footnotesize}
        #1
      \end{footnotesize}
    \end{minipage}
  \end{center}
  \bigskip
}

\begin{document}

%
%
%
%
%
{  

\makeatletter
\@ifundefined{c@affiliation}%
{\newcounter{affiliation}}{}%
\makeatother
\newcommand{\affiliation}[2][]{\setcounter{affiliation}{#2}%
\ensuremath{{^{\alph{affiliation}}}\text{#1}}}
%

\contribution [Application of High Quality Antiproton Beam\\to Study Charmonium and Charmed Hybrids.]  
{Application of High Quality Antiproton Beam\\with Momentum Ranging from 1 GeV/c to 15 GeV/c\\to Study Charmonium and Charmed Hybrids.}  
{M.Yu.}{Barabanov}
{Joint Institute for Nuclear Research,\\
Veksler and Baldin Laboratory of High Energy Physics\\
141980 Dubna, RUSSIA}  
{barabanov@sunhe.jinr.ru}  
{\!\!, A.S. Vodopyanov, V.Kh. Dodokhov and V.A. Babkin} 
%

\abstract{ The elaborate analysis of spectrum of charmonium states
and charmed hybrids in the mass region over $D\bar{D}$-threshold
is given. The combined approach based on the potential model and
relativistic spherical symmetric top model for decay products has
been proposed. The experimental data from different collaborations
were analyzed. Especial attention was given to the new states with
the hidden charm discovered recently. Eight of these states may be
interpreted as higher laying radial excited charmonium states. But
much more data on different decay modes are needed for deeper
analysis. These data can be derived directly from the experiments
using high quality antiproton beam with the momentum ranging from
1~GeV/c to 15~GeV/c (PANDA experiment at FAIR). }
%

\section{Introduction}

The study of charmonium and charmed hybrids spectroscopy is one of
the main domains of elementary particle physics. It seems to be a
challenge nowadays. The research of charmonium (the system
consisting of charmed quark-antiquark pair $c\bar{c}$) and charmed
hybrids (the system consisting of charmed quark-antiquark pair
strongly interacting with gluonic component $c\bar{c}$g) using the
antiproton beam with momentum ranging from 1~GeV/c to 15~GeV/c in
PANDA experiment at FAIR is perspective and interesting from the
scientific point of view. Charmonium and charmed hybrids with
different quantum numbers are copiously produced in
antiproton-proton annihilation process. The accuracy of mass and
width measurements depends only on the quality of antiproton beam
(high luminosity, minimal beam momentum spread, small lateral beam
dimension). It becomes possible to extract the information about
excited states of charmonium which can be extremely useful for
understanding the nature of strong interactions. The performed
analysis of charmonium is promising to understand the dynamics of
quark interactions at small distances~\cite{panda}.

\section{Results of calculations}

The charmonium system has been investigated in detail, first, in
$e^+e^-$-reactions, and afterwards --- on a restricted scale but
with high precision ---  in  $\bar{p}p$-annihilations (the
experiments R704 at CERN and E760/E835 at Fermilab). Nowadays the
scalar $^{1}D_{2}$ and vector $^{3}D_{J}$ charmonium states are
not established. The higher laying scalar $^{1}S_{0}$, $^{1}P_1$
and vector $^{3}S_1$, $^{3}P_J$ charmonium states are
 badly investigated~\cite{RPPh}. The domain over $D\bar{D}$-threshold of 3.73 GeV/c$^2$ is poorly studied.
 According to the contemporary quark models (LQCD, flux tube model), namely in this domain, the
existence of charmed hybrids with exotic ($J^{PC}=0^{+-}, 1^{-+}, 2^{+-}$) as with non-exotic
 ($J^{PC}=0^{-+}, 1^{+-}, 2^{-+}, 1^{++}, 1^{--}$) quantum numbers is expected \cite{panda,RPPh}.

The elaborated analysis of spectrum of the scalar ($^{1}S_0,
^{1}P_1, ^{1}D_2$), vector ($^{3}S_1, ^{3}P_J, ^{3}D_J$)
charmonium states and charmed hybrids with exotic and non-exotic
quantum numbers in the mass region mainly over
$D\bar{D}$-threshold, has been fulfilled~\cite{Bara3, Bara4}.
Different decay modes of charmonium such as decays into
particle-antiparticle or $D\bar{D}$-pair, decays into light
hadrons and decays with $J/\Psi$ in the final state were
investigated. Concerning the charmed
 hybrids, the decays into charmonium and light mesons in the final state and decays into $D\bar{D}^{*}$-pair,
were, in particular, analyzed. These modes possess small widths and significant branching ratios.
This fact facilitates their experimental detection.

   Using the combined approach based on the quarkonium potential model and relativistic top model
 for decay products, ten new radial excited states of charmonium were predicted in the mass region
over $D\bar{D}$-threshold equal to 3.73~GeV/c$^2$. Sixteen charmed
hybrids (lowest-laying hybrids and their radial excited states)
are expected to exist in the discussed mass region.
A special attention is given to the new states with the hidden
charm discovered recently ($XYZ$-particles)~\cite{Eichten,
Brambilla}. The experimental data from different collaborations
(Belle, BaBar, CLEO, CDF) were carefully analyzed. It has been
found that eight of new recently discovered states may be
 interpreted as charmonium states (two scalar $^{1}S_0$, three vector $^{3}S_1$ and tree vector
$^{3}P_J$). But much more data on different decay channels (modes) are needed for deeper analysis.
These data can be derived directly from PANDA experiment with its high quality antiproton beam.
Hence, there is a possibility of measuring the masses, widths and
branching ratios of different charmonium and charmed hybrid states with high accuracy.

   Figure~\ref{fig:spect} illustrates the spectrum of scalar $^{1}S_0$ and vector $^{3}S_1$, $^{3}P_J$ states of
charmonium. Black boxes correspond to the established charmonium states, black-white boxes --- recently
experimentally revealed states with the hidden charm ($XYZ$-particles) that may be interpreted as higher
laying charmonium states. Possible existence of charmonium states marked by black-white boxes was predicted
in our recent calculations. One can find that $X(3940)$ and $X(4160)$ can be interpreted as radial excited
scalar $^{1}S_0$ states of charmonium; $Y(4260)$, $Y(4360)$ and $Y(4660)$ --- as radial excited vector
$^{3}S_1$ states of charmonium and $X(3915)$, $Y(3940)$, $Z(3930)$ --- as radial excited vector $^{3}P_J$
states of charmonium. Finally, white boxes correspond to the states which are not found yet. But a possibility
of existence of these states is predicted in the framework of the combined approach. They may also be
interpreted as higher laying radial excited states of charmonium.

\begin{figure}[htb]
  \begin{center}
    \includegraphics[width=0.49\textwidth]{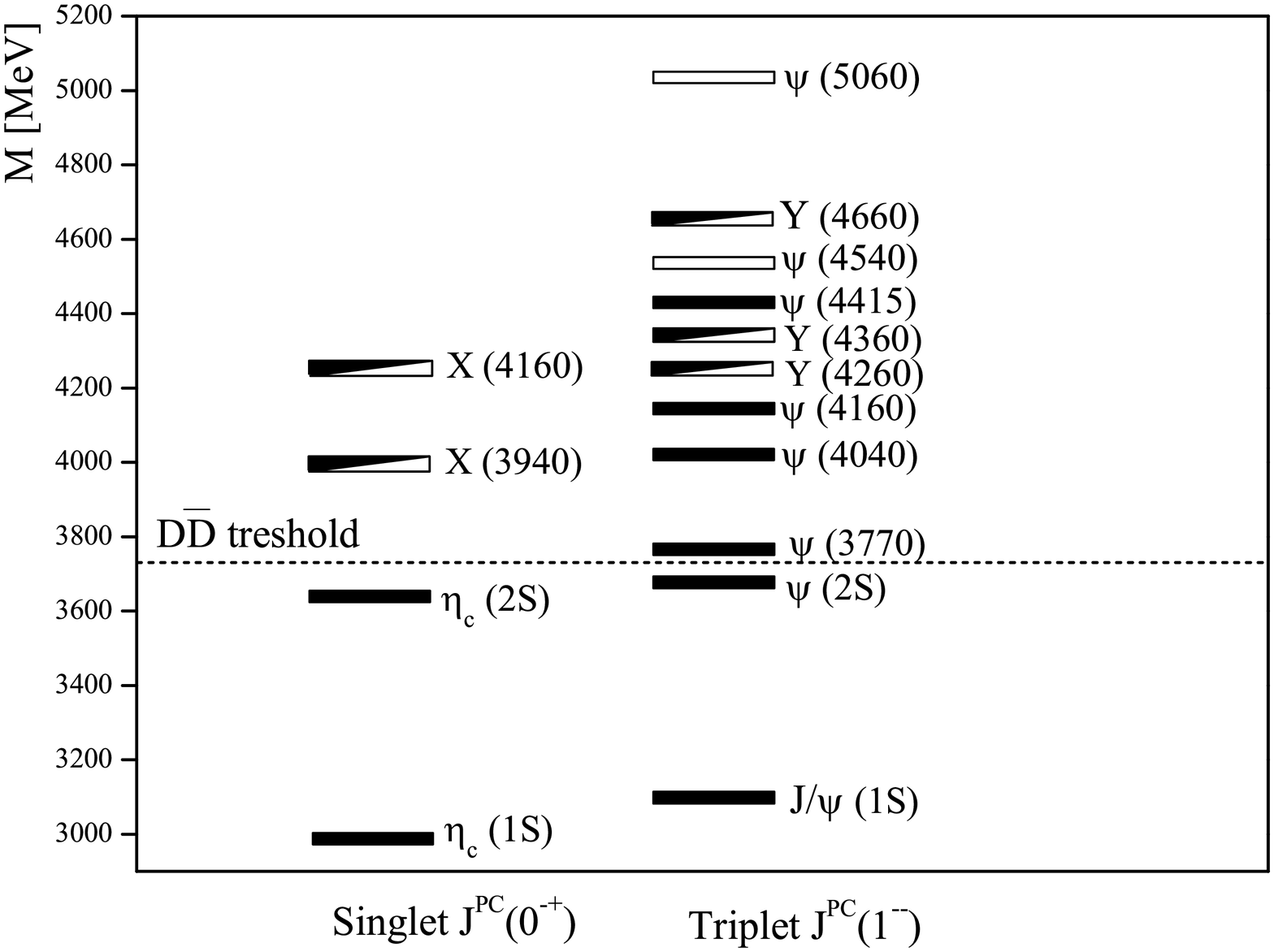}
    \includegraphics[width=0.49\textwidth]{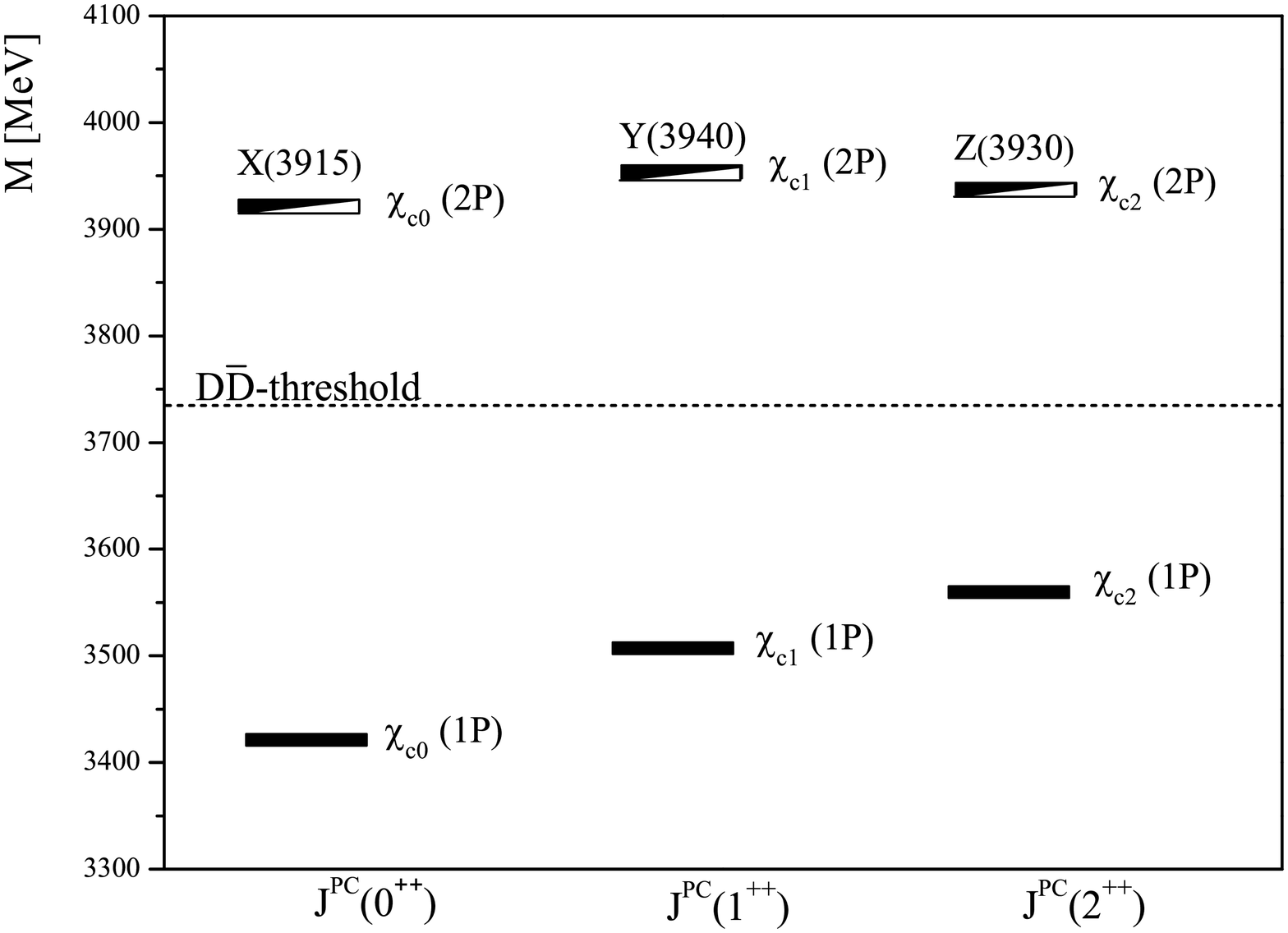}
    \caption{The spectrum of scalar $^{1}S_0$ and vector $^{3}S_1$ and $^{3}P_J$ states of charmonium.}
    \label{fig:spect}
  \end{center}
\end{figure}
%

\begin{figure}[htb]
  \begin{center}
    \includegraphics[width=0.49\textwidth]{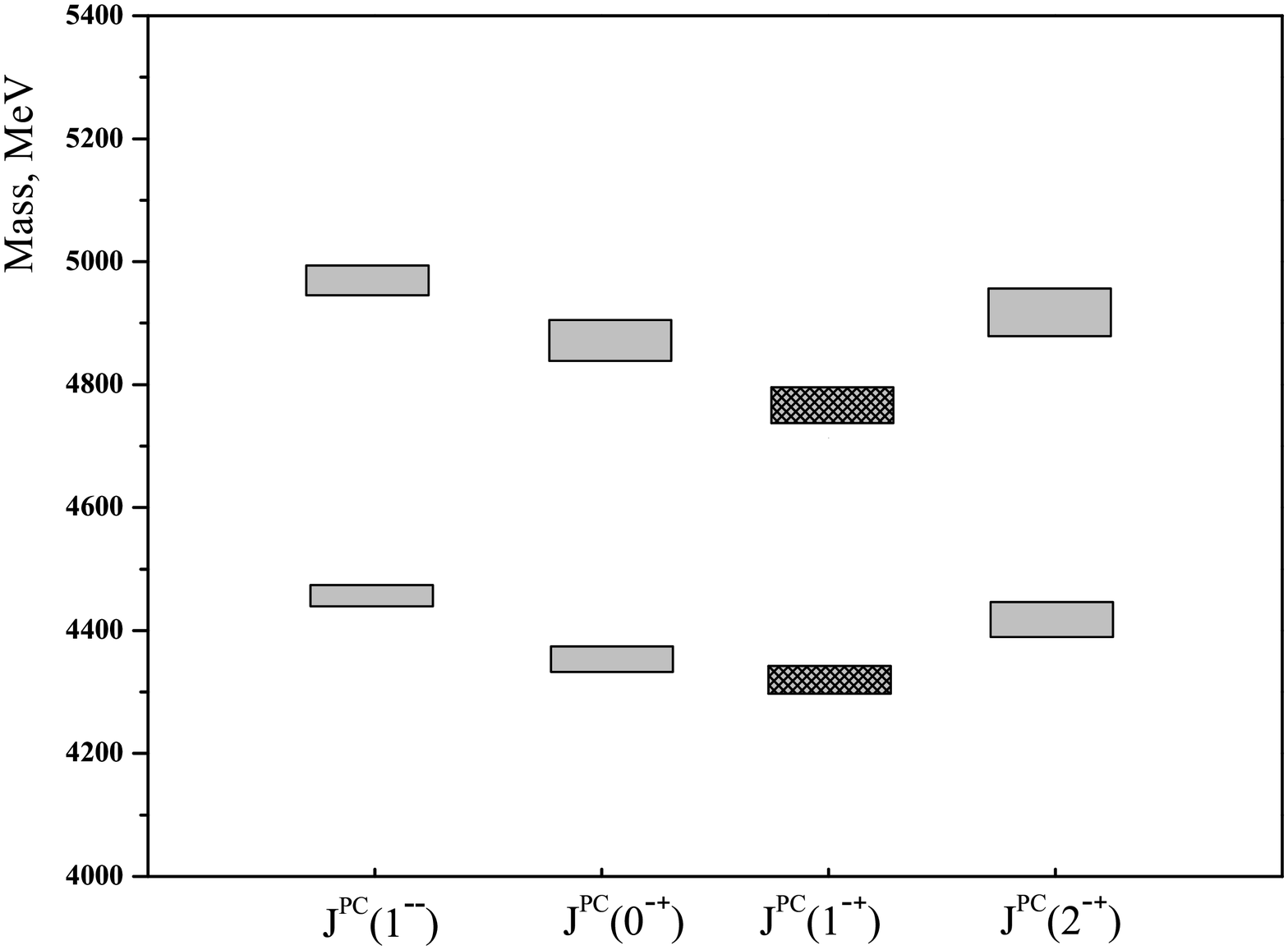}
    \includegraphics[width=0.49\textwidth]{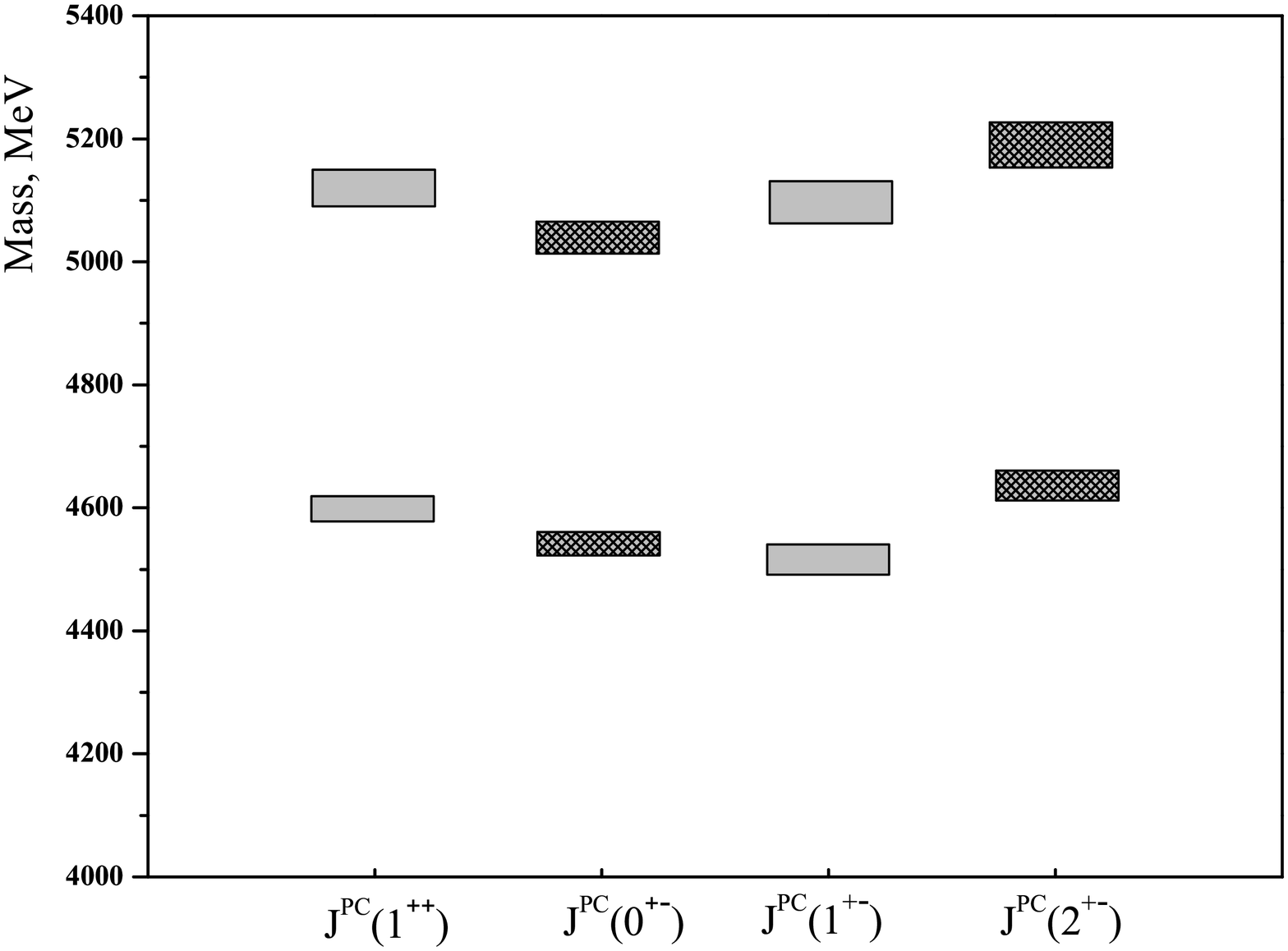}
    \caption{The spectrum of charmed hybrids with quantum numbers $J^{PC}=2^{-+}, 1^{-+}, 1^{--}, 0^{-+}$ and $2^{+-}, 1^{+-}, 1^{++}, 0^{+-}$.}
    \label{fig:hybrid_sp}
  \end{center}
\end{figure}
%

  Figure~\ref{fig:hybrid_sp} illustrates the spectrum of the lowest-laying charmonium hybrids and their radial excited states.
Charmed hybrids with exotic quantum numbers are marked with dark colour and charmed hybrids with
nonexotic quantum numbers --- with light colour. One can find that the state with exotic quantum numbers
$J^{PC}=1^{-+}$ has the lowest mass equals to 4320 MeV/$c^2$. The results of calculations are in good
agreement with the well accepted picture that the quartet $1^{--},(0,1,2)^{-+}$ is lower in mass than
$1^{++},(0,1,2)^{+-}$. The expected splitting is about 150 -- 250 MeV/c$^2$.

   To be sure that the predicted charmonium and charmed hybrid states can really exist and can be found
experimentally, their widths have been calculated~\cite{Bara3,
Bara4}. The values of widths are of an order of several tens of
MeV. This fact facilitates their experimental search.

\section{Conclusions}

Finally the progress of the future charmonium and charmed hybrids researches at FAIR is related to the
results obtained below:

$\bullet$ A combined approach has been proposed to study charmonium and charmed hybrids on the basis
of quarkonium potential model and relativistic top model for decay products.

$\bullet$ Several promising decay channels of charmonium like decays into light hadrons
$\bar{p}{p}\rightarrow{c}\bar{c}\rightarrow\rho\pi$, decays into particle-antiparticle
$\bar{p}{p}\rightarrow{c}\bar{c}\rightarrow\Sigma^{0}\bar{\Sigma} {^0}$, decays into $D\bar{D}$-pair
and decays with $J/\Psi$ in the final state $\bar{p}{p}\rightarrow{c}\bar{c}\rightarrow{J}/\Psi + X$,
were, in particular, analyzed.

$\bullet$ Ten radial excited states of charmonium (two scalar $^{1}S_0$, three vector $^{3}S_1$ and
tree vector $^{3}P_J$) above $D\bar{D}$-threshold have been predicted in the framework of the
combined approach.

$\bullet$ Several promising decay channels of the charmed hybrids
like decays into charmonium and light mesons in the final state
$\bar{p}{p}\rightarrow{c}\bar{c}g\rightarrow\chi_{c0,1,2} (\eta,
\pi\pi;...)$, $\bar{p}{p}\rightarrow{c}\bar{c}g\rightarrow{J}/\Psi
(\eta, \omega, \pi\pi;...)$ and decays into $D\bar{D}^*$-pair
$\bar{p}{p}\rightarrow{c}\bar{c}g\rightarrow{D}\bar{D}^*\eta$ were
considered.

$\bullet$ Sixteen charmed hybrids with exotic and nonexotic
quantum numbers are expected to exist in the framework of the
combined approach.

$\bullet$ The recently discovered $XYZ$-particles have been
analyzed. Some of these states can be interpreted as higher laying
radial excited states of charmonium. The necessity of further
studying the $XYZ$-particles and their main characteristics in
PANDA experiment at FAIR has been demonstrated.



This work was supported by the FAIR- Russia Research Center (FRRC).


%

}  



\begin{thebibliography}{99}

\bibitem{panda}
  PANDA Collaboration, Physics Perfomance Report, 63 (2009).

\bibitem{RPPh}
  Review of Particle Physics, Journal of Physics G: Nuclear and Particle Physics, {\bf V. 37}, N. 7A, 1040 (2010).



\bibitem{Bara3}
  M.Yu. Barabanov et al., Hadronic Journal, {\bf V. 32}, N. 2, 159
  (2009).

\bibitem{Bara4}
  M.Yu. Barabanov et al., Proc. of the XX International Seminar on High Energy Physics Problems, Dubna, Russia, Oct 4-9, 137 (2010).



\bibitem{Eichten}
  E. Eichten, S. Godfrey, J. Rosner, Reviews of Modern Physics, {\bf V. 80},  N. 3, 1161 (2008).

\bibitem{Brambilla}
  N. Brambilla et al., European Physical Journal, C 71 :1534, 1 (2011).



\end{thebibliography}
\end{document}